\documentclass[
showpacs,superscriptaddress,twocolumn, aps, longbibliography]{revtex4-1} 

\usepackage[USenglish]{babel}

\usepackage{graphicx,xcolor}
\usepackage{amsmath, bbm}
\usepackage{mathtools}
\usepackage{natbib}
\usepackage{hyperref}
\usepackage{epstopdf}
\usepackage{color}

\renewcommand{\imath}[0]{\mathrm{i}}

\newcommand{\myref}[1]{}

\begin{document}
\title{Coherent coupling of a single molecule to a \\scanning Fabry-Perot microcavity}
 \author{Daqing Wang}%
\affiliation{Max Planck Institute for the Science of Light, D-91058 Erlangen, Germany}
 \author{Hrishikesh Kelkar}%
\affiliation{Max Planck Institute for the Science of Light, D-91058 Erlangen, Germany}
\author{Diego Martin-Cano}%
\affiliation{Max Planck Institute for the Science of Light, D-91058 Erlangen, Germany}
\author{Tobias Utikal}%
\affiliation{Max Planck Institute for the Science of Light, D-91058 Erlangen, Germany}
\author{Stephan G\"otzinger}%
\affiliation{Friedrich Alexander University Erlangen-Nuremberg, D-91058 Erlangen, Germany}
\affiliation{Max Planck Institute for the Science of Light, D-91058 Erlangen, Germany}
\affiliation{Graduate School in Advanced Optical Technologies (SAOT), Friedrich Alexander
University Erlangen-Nuremberg, D-91052 Erlangen, Germany.}
\author{Vahid Sandoghdar}
\affiliation{Max Planck Institute for the Science of Light, D-91058 Erlangen, Germany}
\affiliation{Friedrich Alexander University Erlangen-Nuremberg, D-91058 Erlangen, Germany}

\date{\today}

\begin{abstract}
Organic dye molecules have been used in great many scientific and technological applications, but their wider use in quantum optics has been hampered by transitions to short-lived vibrational levels, which limit their coherence properties. To remedy this, one can take advantage of optical resonators. Here we present the first results on coherent molecule-resonator coupling, where a single polycyclic aromatic hydrocarbon molecule extinguishes 38\% of the light entering a microcavity at liquid helium temperature. We also demonstrate four-fold improvement of single-molecule stimulated emission compared to free-space focusing and set the ground for coherent mechanical manipulation of the molecular transition. Our experimental approach based on a microcavity of low mode volume and low quality factor paves the way for the realization of various nonlinear and collective quantum optical effects with molecules.

\end{abstract} 

\maketitle

\section{Introduction}
Organic dye molecules are omnipresent in our everyday life and have had a large impact on optical sciences. One of their most well-known applications is in dye lasers \cite{Duarte-book}, which pioneered laser technology and continue to offer laser properties that are not easily  attainable from other media. A further important contribution of dye molecules has been in fluorescence microscopy and the advent of single-molecule spectroscopy, which in turn led to the optical detection of other quantum emitters in the solid state such as semiconductor quantum dots and color centers \cite{Moerner:99}. More recently, the rich photophysics of dye molecules was exploited to invent super-resolution optical microscopy \cite{Weisenburger:15}. 

Since the early 1990's, a number of experiments have shown that a particular class of dye molecules, the polycyclic aromatic hydrocarbons (PAH), behave favorably as quantum emitters when embedded in organic crystals at low temperatures, $T<4$\,K \cite{SMbook, Tamarat:95, Kulzer:97, Lounis:97, Brunel:98, Brunel:99, Hettich:02, Gerhardt:07a, Wrigge:08, Hwang:09, Lettow:10, Pototschnig:11, Rezus:12, Faez:14, Siyushev:14, Maser:16}. Resembling cut-out pieces of a graphene sheet, PAHs can be designed and chemically synthesized in several configurations and transition wavelengths \cite{Harvey-book}. In particular, PAH systems can be operated at the transition wavelengths of other quantum emitter species such as Na, Li, Rb, or $\rm Pr^{3+}$, making them ideal for use in hybrid quantum platforms \cite{Siyushev:14}. 

PAHs can serve as solid-state emitters with remarkable optical properties such as near-unity quantum efficiency and indefinite photostability. In particular, at low temperatures they also support natural-linewidth-limited zero-phonon lines (00ZPL) between the ground vibrational levels of the electronic ground ($\left|{g, v=0}\right>$) and excited ($\left|{e, v=0}\right>$) states (see Jablonski diagram in inset of Fig.\,\ref{cavity}b). These features make it possible to couple a propagating laser beam or a stream of single photons to a single molecule with high efficiency, e.g., via tight focusing or dielectric nanoguides \cite{Gerhardt:07a, Wrigge:08, Pototschnig:11, Rezus:12, Faez:14}. In addition, this efficient coupling also opens the door to nonlinear optical effects such as stimulated emission, three-photon amplification and four-wave mixing in a single molecule with only a few photons \cite{Hwang:09, Maser:16}. However, transitions to the higher vibrational levels of the electronic ground state (Frank-Condon principle) and a small coupling to the host matrix vibrations (Debye-Waller factor) reduce the scattering cross section by a branching ratio (fraction of emission on the 00ZPL to the total decay of the excited state) of about $\alpha$=30-50\% from the ideal value of $3\lambda^2/2\pi$. As a result, a perfect coupling between a propagating photon and a molecule \cite{Zumofen:08} remains out of reach. To render the molecule as a two-level atom and compensate for this shortcoming, one can enhance the 00ZPL emission by using the Purcell effect \cite{Purcell1946} and obtain a near-unity branching ratio. In this work, we report the first cavity-enhanced coherent coupling of a single molecule to a light field, demonstrate its potential for nonlinear switching, and discuss future prospects for achieving full coupling.

\begin{figure}[t]
\centering 
 \includegraphics[width=0.9\columnwidth]{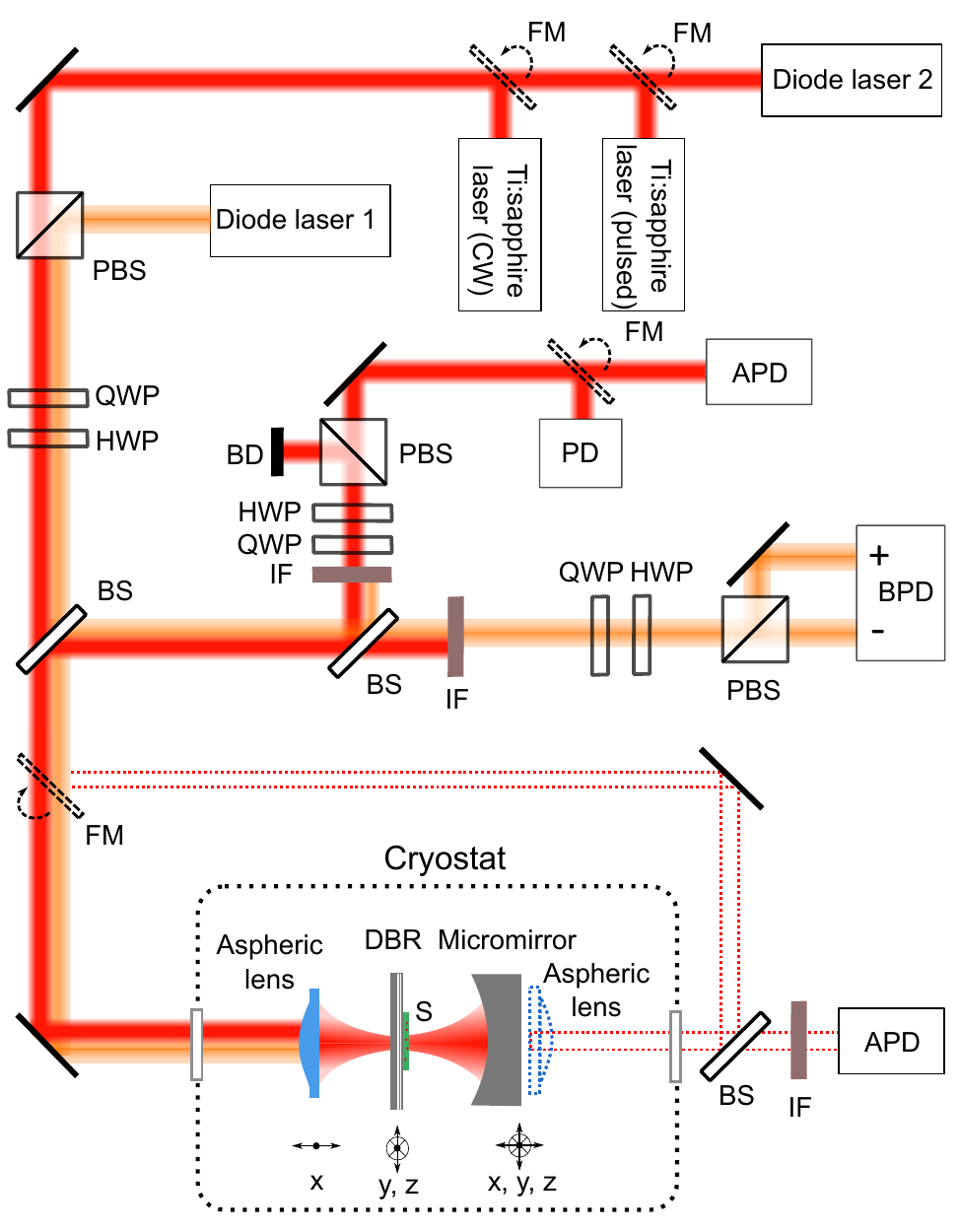}
 \caption{Schematics of the experimental arrangement. Slip-stick piezo-electric motors and transducers were used to control the position of the micromirror both in the lateral and  axial directions. An aspheric lens with a numerical aperture NA=0.55 was employed to couple an incoming laser beam to the microcavity. The dashed lines and elements signify a second aspheric lens and the laser beams used to access the sample from the other side in the absence of the micromirror. S: sample, PD: photodiode, BPD: balanced photodetector, APD: avalanche photodiode, BS: beam splitter, PBS: polarizing beam splitter, HWP: half-wave plate, QWP: quarter-wave plate, IF: interference filter, BD: beam dump, FM: flip mirror. See text for details. }
\label{schematics}
\end{figure}

\section{Experimental matters}
\begin{figure}
\centering 
 \includegraphics[width=0.9\columnwidth]{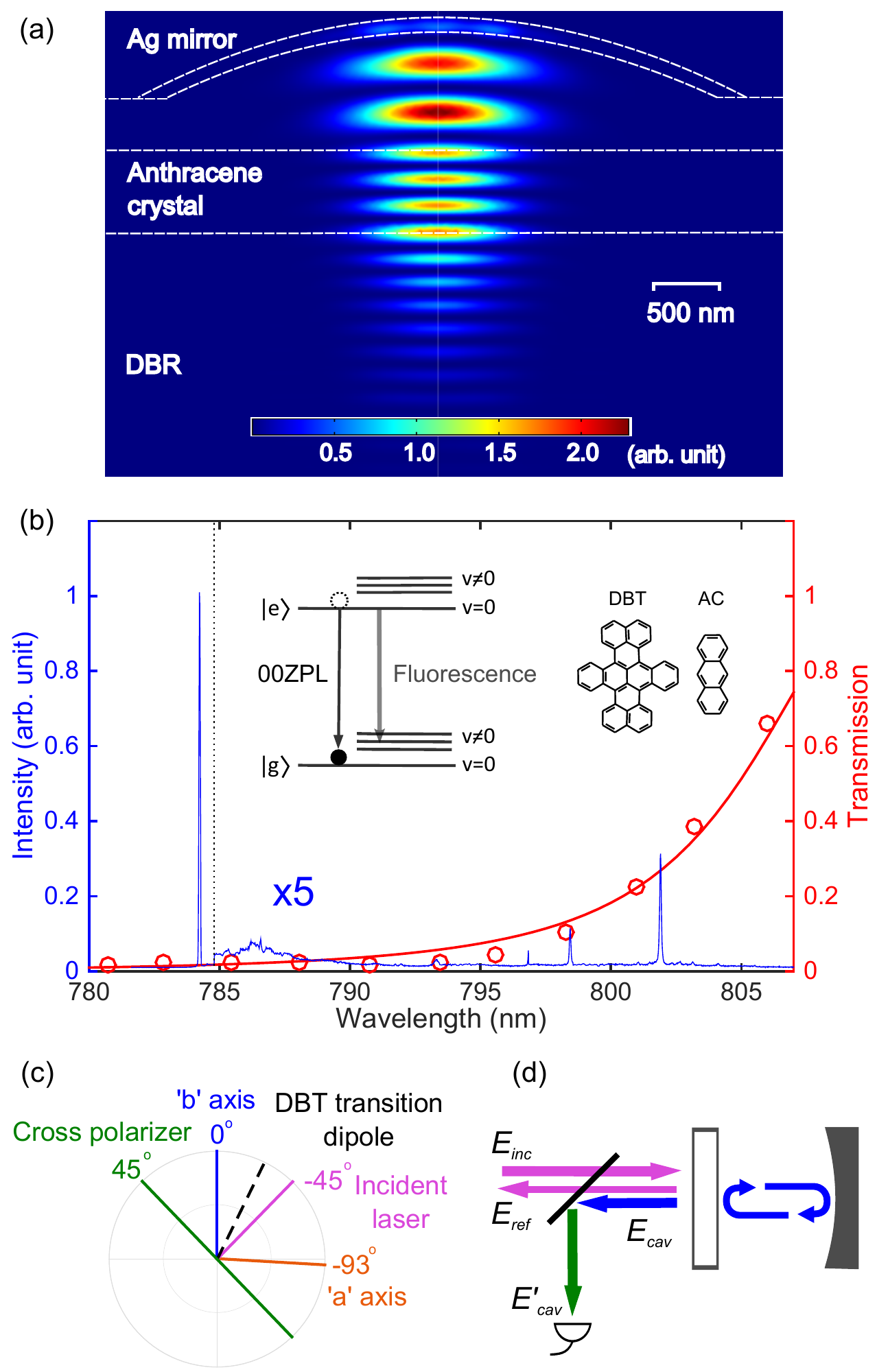}
 \caption{a) Microcavity design to scale, showing the electric field intensity distributions in a cavity mode. b) The red curves show the measured (symbols) and calculated (solid line) transmission of the DBR at liquid helium temperature as a function of wavelength. The blue spectrum presents the emission of a molecule excited at 766\,nm. The section beyond 785 nm is magnified by five times for clarity. c) Relative position of AC crystal axes, laser and detection polarizations as well as the orientation of the DBT dipole moment in the DBR plane. d) The arrangement for cross polarized measurements.}
\label{cavity}
\end{figure}

Figure \ref{schematics} shows the overall schematics of our experiment. At the core of the setup, a microcavity is formed between a planar distributed Bragg reflector (DBR) and a micromirror with a radius of curvature of 5\,$\mu$m and depth of 550\,nm (also see Fig.\,\ref{cavity}a). The DBR was designed with reflectivity $R =99.5\%$ for a plane wave at wavelength $\lambda=780$\,nm, while the micromirror was produced by focused ion beam milling of a silicon wafer followed by a protected silver coating with a nominal reflectivity $R =99\%$. Being fabricated on a microscopic pedestal, the curved micromirror can be positioned and scanned in all three dimensions at will. The fabrication and characterization details of such a cavity can be found in Ref. \cite{Kelkar:15}. A thin (600\,nm) crystal of anthracene (AC) doped with dibenzoterrylene molecules (DBT, see inset in Fig.\,\ref{cavity}b) was produced in a co-sublimation chamber and was placed on the DBR surface (see Fig.\,\ref{cavity}a). By monitoring the reflection of an incident laser beam from the cavity and varying its length over one free spectral range, we deduced a cavity finesse of $\cal{F}=$200. 

The red symbols in Fig. \ref{cavity}b show the band edge of the DBR measured in transmission when we replaced the micromirror by an aspheric lens (see Fig.\,\ref{schematics}). The outcome agrees well with the design calculations shown by the solid red curve. The blue spectrum displays the emission of a DBT molecule upon excitation from the vibrational ground level of the electronic ground state ($\left|{g, v=0}\right>$) to a higher vibrational level of the electronic excited state ($\left|{e, v\not=0}\right>$) at wavelength $\lambda=766$\,nm. This signal was recorded by exciting the sample through the aspheric lens that replaced the micromirror (see Fig.\,\ref{schematics}). The strong sharp peak denotes the 00ZPL, while the small shoulder next to it represents the phonon wing caused by the coupling to the lattice vibrations of the AC matrix. The red-shifted weak resonances originate from the fluorescence of $\left|{e, v=0}\right>$ to higher-order vibrational levels of the electronic ground state $\left|{g, v\not=0}\right>$. 

The intensity distribution in Fig.\,\ref{cavity}a displays the calculated profile of the bare cavity mode with volume $V\sim1.7\,\lambda^3$. To access the field inside the cavity, we take advantage of the cavity birefringence caused by the AC crystal with two orthogonal axes \textbf{a} and \textbf{b} in the DBR plane\,\cite{Nakada:62}. As illustrated in Fig.\,\ref{cavity}c, our measurements yielded an angular separation of $93^\circ$ for the projections of these two axes. Considering that the dipole moments of DBT molecules can be oriented at angles between zero and about $30^\circ$ \cite{nicolet2007b, Polisseni:16}, we aligned the polarization of the incident laser beam at $-45^\circ$ from \textbf{b} and placed a cross polarizer at $+45^\circ$. As illustrated in Fig.\,\ref{cavity}d, a part of the incident beam ($\rm E_{inc}$, pink) is coupled into the cavity ($\rm E_{cav}$, blue), exits it and traverses the cross polarizer with 50\% efficiency ($\rm E'_{cav}$, green). The field $\rm E'_{cav}$ detected on a very low background provides the means for monitoring the field inside the cavity, equivalent to measuring the transmission of a symmetric resonator.

A mirror displacement of about 1\,nm is sufficient to detune the resonance of a cavity with finesse $\cal{F}=$200. While the helium bath cryostat helps to achieve a stable temperature, nanometer vibrations are very difficult to eliminate in a passive manner. To minimize mechanical perturbations, we performed most of our measurements at temperatures $T\sim3-3.6\,K$ to avoid the need for strong pumping on the liquid helium tank. In addition, we used a H\"ansch-Couillaud scheme \cite{Haensch:80} to lock the cavity resonance to a narrow-band diode laser at $\lambda=762$\,nm. Here, we sent a diode laser beam with polarization aligned to the cross polarizer (see Figs.\,\ref{schematics}, \ref{cavity}c) and frequency tuned to a cavity resonance along the \textbf{a} axis. A dispersive error signal created by a balanced photodetector allows us to achieve a stable cavity lock at root-mean-square displacement of 0.1 nm.  

\begin{figure}
\centering 
 \includegraphics[width=\columnwidth]{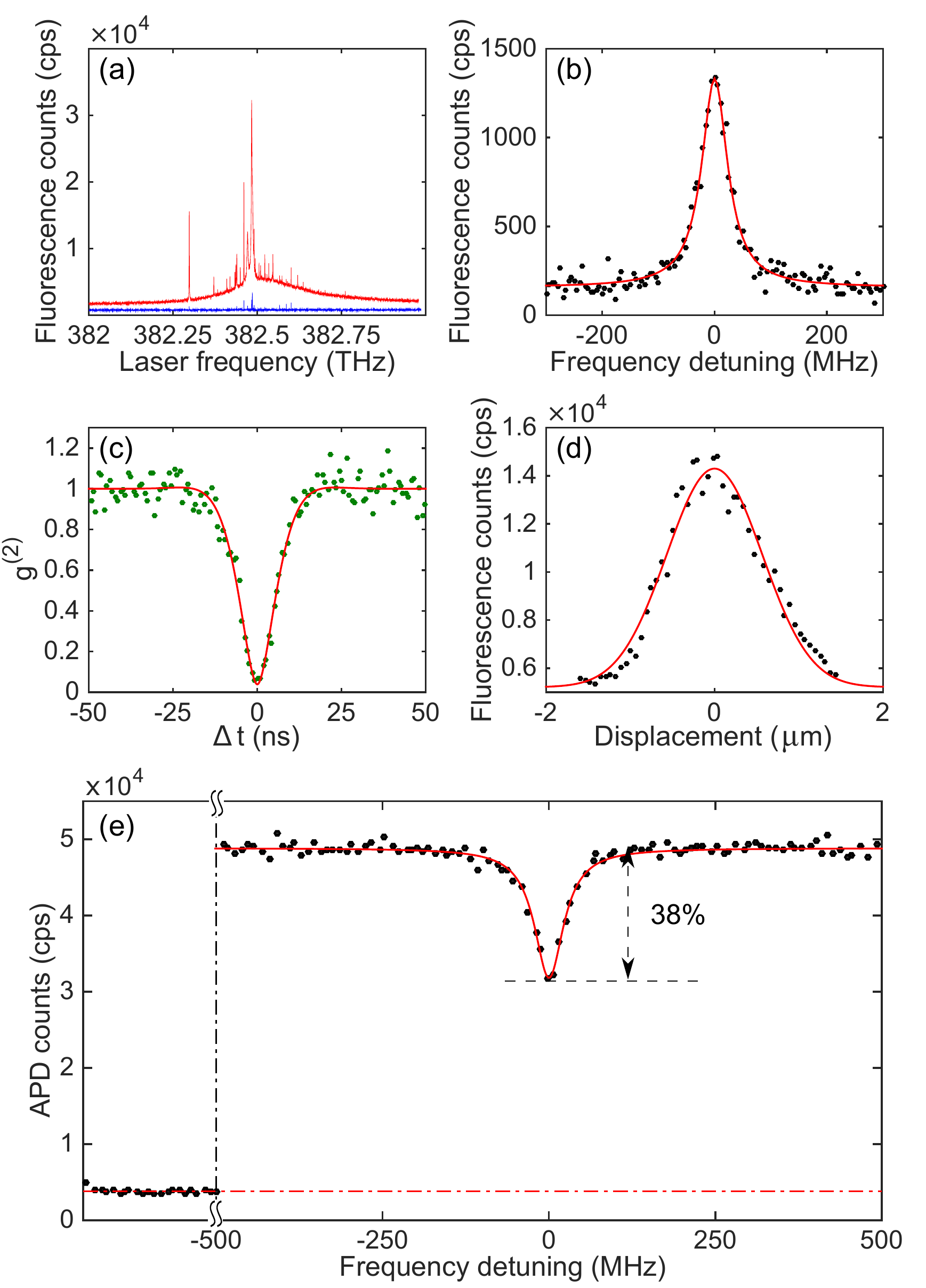}
 \caption{a) The red and blue spectra show the detected Stokes shifted fluorescence for $\lambda>785$\,nm when an incident laser was scanned across the inhomogeneous broadening of the molecules for the cases of cavity locked on resonance and micromirror retracted, respectively. The width of the red envelope gives a measure for the cavity FWHM of 250 GHz. b) A zoom of one sharp resonance with a Lorentzian fit. c) $g^{(2)}$ measured via a coincidence Hanbury-Brown and Twiss study. d) A lateral cross section of the cavity mode in the AC crystal mapped by measuring the fluorescence of a single molecule. The solid curve shows a Gaussian fit. e) Response of the resonant cavity-molecule system as a function of laser frequency detuning. The left side shows the signal for the laser frequency far from the cavity resonance. }
\label{fluorescence-excitation}
\end{figure}

\section{Results}
Having discussed the basic arrangement of the microcavity, in Fig.\,\ref{fluorescence-excitation} we begin to present our studies on coupling it to single DBT molecules. The red curve in Fig.\,\ref{fluorescence-excitation}a shows the Stokes-shifted fluorescence ($\left|{e, v=0}\right>\rightarrow \left|{g, v\not=0}\right>$) of the sample as a the laser frequency was scanned through the 00ZPLs of the molecules within the cavity resonance. This signal leaks out of the DBR because the latter was designed to have a much lower reflectivity in the Stokes-shifted spectral regime (see Fig.\,\ref{cavity}b). The sharp resonances in the red curve correspond to different individual molecules within the sample inhomogeneous band of about 500\,GHz. The envelope stems from the weak fluorescence of the background molecules in the excitation spot and reveals the cavity linewidth, which allows us to deduce a quality factor of $Q$=1500. 

Figure\,\ref{fluorescence-excitation}b shows a zoom onto one of the single-molecule 00ZPL resonances at $\lambda \sim 784.3$\,nm. To verify that this represents a single molecule, we recorded the second-order autocorrelation function $g^{(2)}$, plotted in Fig.\,\ref{fluorescence-excitation}c. The obtained value of $g^{(2)}(0)=0.04$ confirms a high-purity single-photon source.  We remark that the distribution of the resonance heights in Fig.\,\ref{fluorescence-excitation}a does not directly correlate with the cavity resonance profile. This is because the observed fluorescence intensity is also a function of the excitation strength of the individual molecules, which is influenced by their lateral positions within the cavity mode. Figure\,\ref{fluorescence-excitation}d displays a lateral cross section of the spatial cavity mode mapped by the fluorescence of a single molecule as the DBR was scanned laterally. This allows us to put an upper bound of $1.3\,\mu$m on the full width at half-maximum (FWHM) of the cavity mode, considering that the molecule might be saturated at the spot center. 

The blue curve in Fig.\,\ref{fluorescence-excitation}a plots the signal when the micromirror was retracted. A comparison of the blue and red curves in this figure emphasizes the low transmissivity of  the DBR for the excitation signal and the cavity enhancement of the fluorescence signal. To estimate the effect of the cavity on the branching ratio, we deduced the excited state lifetime of the same molecule when resonantly coupled to the full cavity ($\tau^\text{(cav)}=3.2\pm 0.1$\,ns) and in the absence of the micromirror ($\tau^\text{(DBR)}=3.9\pm 0.1$\,ns) from Hanbury-Brown and Twiss measurements. The resulting values match the measured linewidths of $\gamma^\text{(cav)}/2\pi\approx50$\,MHz and $\gamma^\text{(DBR)}/2\pi\approx40$\,MHz. Considering a branching ratio $\alpha^{\rm (AC)}$=0.33 for DBT in anthracene \cite{Trebbia:09}, the observed lifetime reduction implies 66\% enhancement of the 00ZPL transition and a modified branching ratio of $\alpha^\text{(cav)}=0.45$ in the cavity. 

Next, we examine the cavity effect on the coherent response of the molecule. Figure\,\ref{fluorescence-excitation}e shows the intensity of the excitation laser beam after reflection from the cavity and through the cross polarizer. A narrow Lorentzian profile with FWHM of $\gamma^{\rm (cav)}$=54\,MHz is observed when the laser frequency scans through the 00ZPL of a single molecule. A transmission dip of 38\% provides a direct measure for the interaction efficiency of a single molecule with a photon inside the cavity and can be seen as a strong dipole-induced reflection of a laser beam. It is important to note that this sharp feature sits on a cavity resonance that is several thousand times broader. The left-side of the intensity trace in Fig.\,\ref{fluorescence-excitation}e displays the signal when the laser frequency was detuned far out of the cavity resonance, defining a reference level for our measurement. 
\begin{figure}
\centering 
\includegraphics[width=\columnwidth]{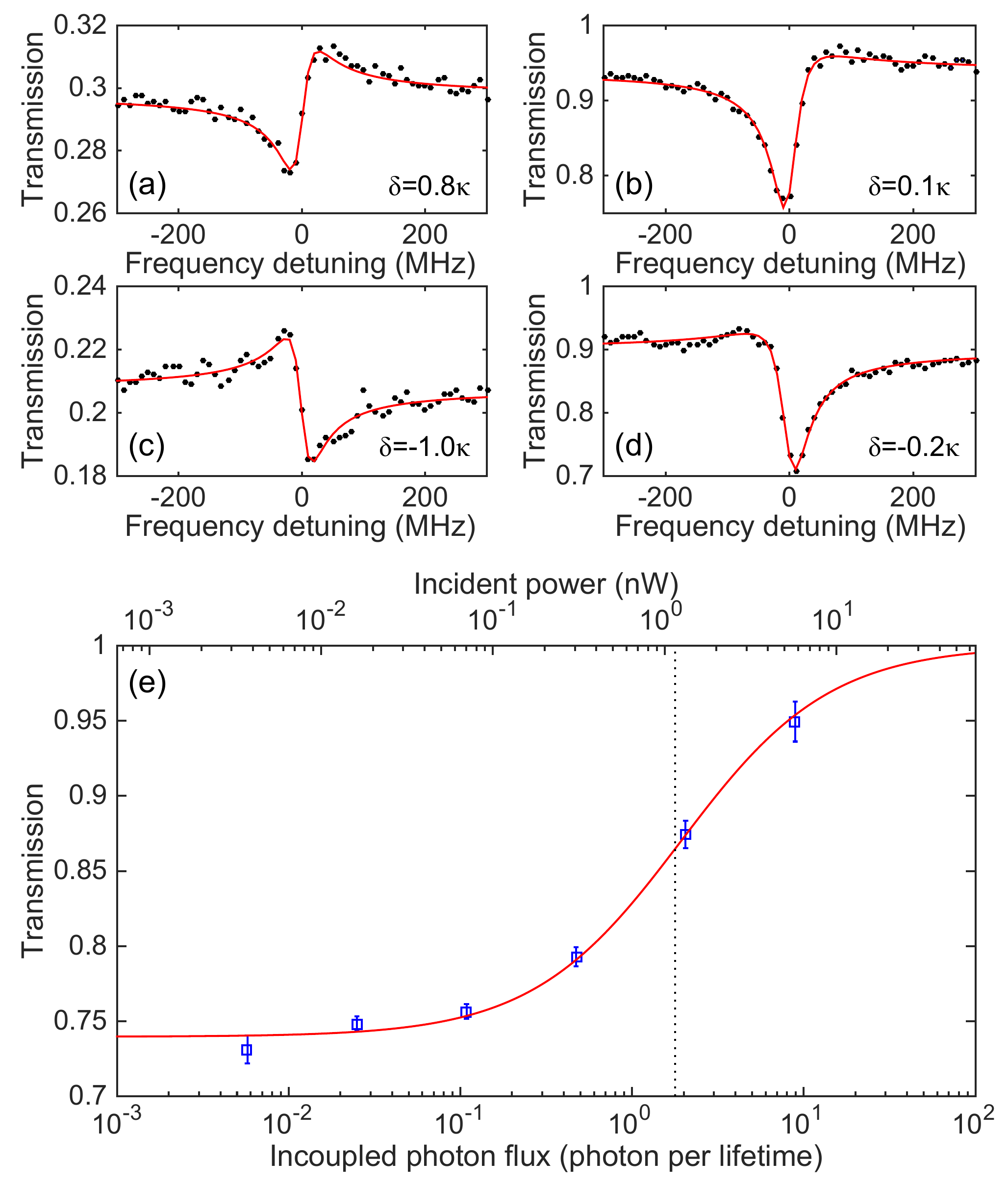}
 \caption{a-d) Response of the resonant cavity-molecule system as a function of laser frequency detuning for four different cavity detunings by the amounts shown in each legend. e) The response of the system as a function of the excitation power. The dotted vertical line indicates saturation parameter $S=1$. In these measurements the finesse had dropped to about one hundred due to condensation of residual gases.}
\label{coherent}
\end{figure}

A paramount feature of coherent interactions between oscillators of different linewidths is a dispersive Fano resonance \cite{Fano:61}. The data in Fig.\,\ref{coherent}a-d show the signals that were obtained when the cavity resonance was detuned with respect to the molecular resonance by an amount comparable to the cavity FWHM, $\kappa$. Figures\,\ref{coherent}a and \ref{coherent}c present clear dispersive resonances when the cavity center frequency was shifted by $\delta=0.8\kappa$ and $\delta=-\kappa$, respectively. The dispersive signals stem from the coherent coupling between the broad resonance of the cavity and the narrow resonance of the molecule \cite{Auffeves:07}. As expected, the spectra approach an absorptive shape when the cavity detuning is reduced (see Fig.\,\ref{coherent}b, d). 

To characterize the transition from coherent to incoherent interactions, in Fig.\,\ref{coherent}e we show the resonant transmission signal of the cavity as a function of power in the incident laser beam. The upper x-axis shows this quantity measured before the cryostat windows, while the lower horizontal axis represents the photon flux coupled in the cavity in units of photon number per lifetime. The red curve presents a theoretical fit using the formula
\begin{equation}
T=\left[ 1-\frac{\beta\alpha^\text{(cav)}}{1+S}\right]^2,
\end{equation}
where $S$ is the saturation parameter and $\beta$ denotes the fraction of the dipolar emission into the cavity mode. We find a critical photon number of 1.8 for reaching $S=1$. Taking into account an extinction value of 38\% measured in Fig.\,\ref{fluorescence-excitation}e for $S\ll1$, we find $\alpha^\text{(cav)}\beta=0.2$ and $\beta=0.47$, which is a further evidence of an efficient coherent coupling between the molecule and the microcavity. The cooperativity factor can be estimated to be $C=4g^2/\kappa\gamma\sim$\,0.22, and the cavity parameters yield $\{g, \kappa, \gamma\}/2\pi=\{740\,\text{MHz}, 250\,\text{GHz}, 40\,\text{MHz}\}$, where $g$ represents the cavity-molecule coupling strength. 

\begin{figure}
\centering 
 \includegraphics[width=\columnwidth]{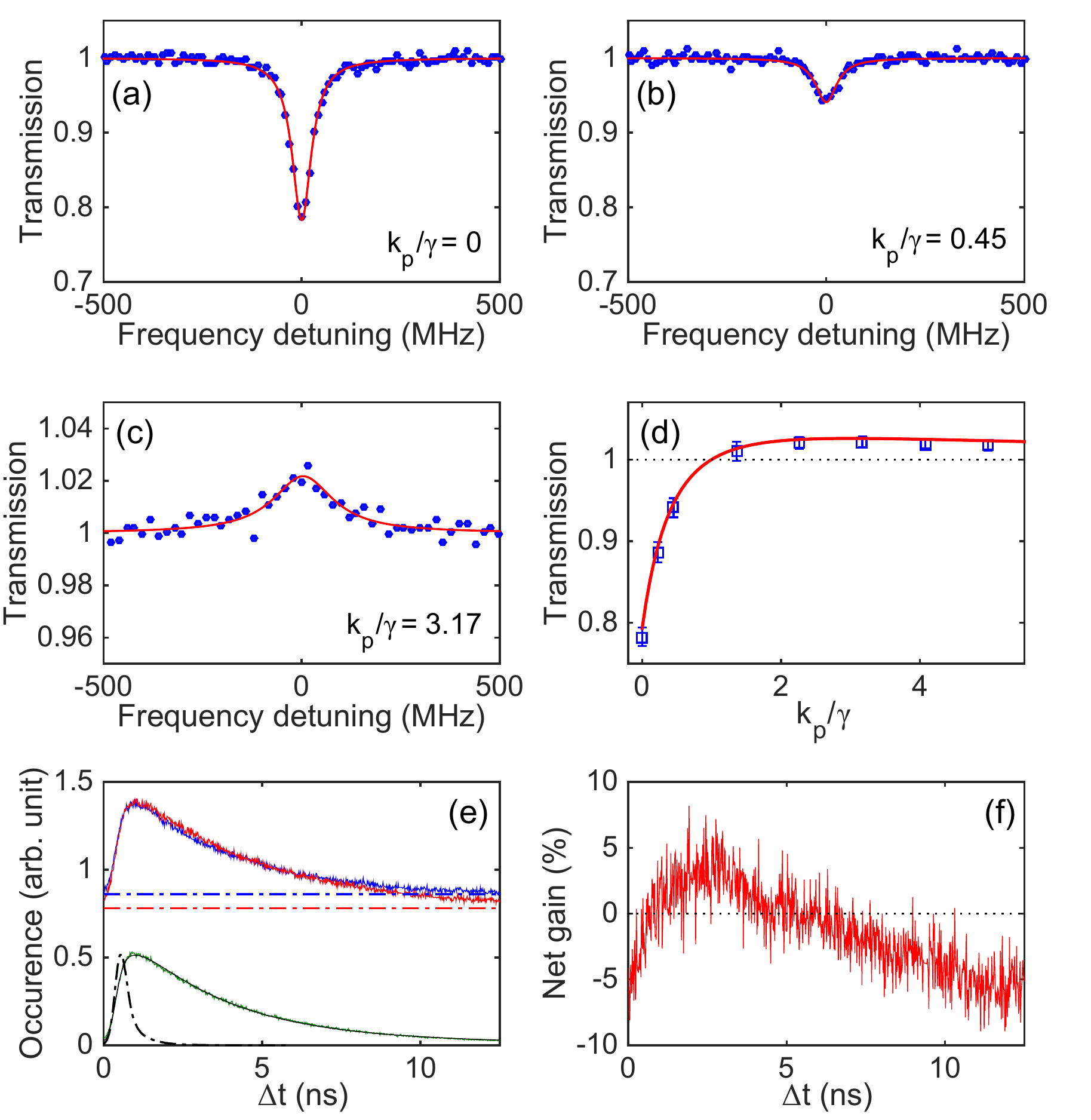}
 \caption{a-c) Coherent resonant response of the cavity-molecule system as a function of the probe laser frequency. In each case, the pump rate $k_{\rm p}$ is indicated in the legend. d) Amplification as a function of pump power. e) The green curve shows the emission of a molecule in response to a picosecond pulse in the absence of the probe laser. The black curve presents a fit, taking into account the instrumental response displayed by the dashed-dotted black profile. The red and blue curves show the time evolution of the emission on the 00ZPL when the probe laser was on and away from the molecular resonance, respectively. The dashed-dotted red and blue horizontal lines mark the background from the probe laser light for the on and off-resonance cases, respectively. f) The net change due to stimulated emission obtained by subtracting the red and blue curves in (e).}
\label{amplification}
\end{figure}

One of the most promising mechanisms for performing photonic operations such as switching is based on the intrinsic nonlinearity of a quantum system \cite{Chang:14}. Recently, it was shown that a single molecule can amplify a weak laser beam that is tightly focused on it by about 0.5\% \cite{Hwang:09}. In Fig.\,\ref{amplification}, we show that the coupling of a molecule to a microcavity can enhance this effect substantially. To do this, we pumped the transition from $\left|{g, v=0}\right>$ to $\left|{e, v\not=0}\right>$ with a continuous-wave Ti:sapphire laser at $\lambda=766$\,nm and recorded the molecule-cavity response to a narrow-band probe laser beam on the 00ZPL resonance of a single molecule. Figure \ref{amplification}a-c displays examples of the evolution of this signal as the pump power was increased and the molecular population was inverted. The symbols in Fig.\,\ref{amplification}d show the measured amplification as a function of the pumping rate, while the solid curve presents a theoretical fit obtained by solving the optical Bloch equations. The maximum continuous-wave amplification of 2\% is about four times larger than the earlier work on free-space focusing on single molecules \cite{Hwang:09}. 

To investigate the temporal behavior of our single-molecule switch, we employed a picosecond Ti:sapphire laser to pump the molecule and monitored the emission of the molecule on its 00ZPL through the cavity as a function of time with respect to the pump pulse. The green curve in Fig.\,\ref{amplification}e shows the evolution of the molecular emission in the absence of the probe laser. A theoretical fit (solid black curve) based on the convolution of the detection instrumental response (black dashed-dotted profile) and an exponential decay of the excited state population lets us deduce a $1/e$ decay time of 3.1\,ns. Next, in Fig.\,\ref{amplification}e we present the buildup and decay of the 00ZPL signal by the red and blue curves for the cases when the probe laser was tuned on and off the 00ZPL, respectively, i.e, depending on whether the molecular emission was stimulated or not. The dashed-dotted red and blue lines in Fig.\,\ref{amplification}e mark the background level of the probe laser for the on and off-resonance cases, respectively; the offset is due to the attenuation caused by the molecule. In Fig.\,\ref{amplification}f, we display the difference of the blue and red curves to visualize the dynamics of population inversion and the evolution from attenuation to gain and back after each pump pulse. The dotted horizontal line marks the transparency of the cavity-molecule system.

\begin{figure}
\centering 
 \includegraphics[width=\columnwidth]{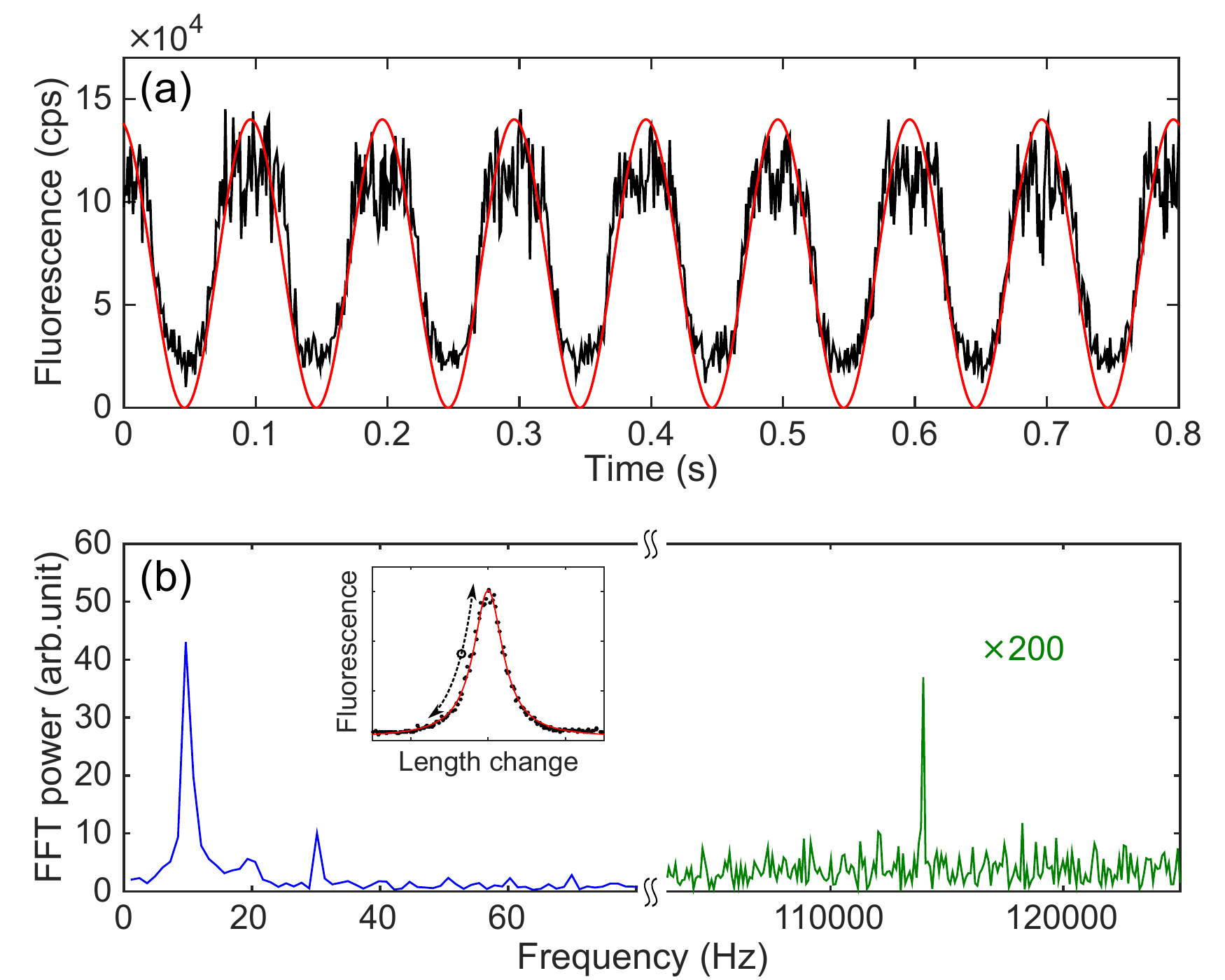}
 \caption{a) 00ZPL emission of a molecule as a function of the cavity length modulation at 10Hz. b) Fourier transform of the signal in (a), revealing the second and third harmonics of it (left) together with the Fourier transform of the signal from a cavity modulation at 114 kHz (right). The inset shows the region of the cavity resonance that is scanned.  The highest frequency achieved in this experiment was limited by the resolution of our data acquisition card.}
\label{mechanics}
\end{figure}

An interesting feature of an open Fabry-Perot cavity is that one can easily manipulate its influence by nanoscopic displacement of its mirrors. It is, thus, possible to actuate the coherent coupling of an incoming photon to the molecule-cavity composite via mechanical interactions. As a first step to exploring this effect, we excited a molecule on a $\left|{g, v=0}\right>$ to $\left|{e, v\not=0}\right>$ transition at $\lambda=766$\,nm and recorded its narrow-band emission on the 00ZPL through a resonant cavity.  In Fig.\,\ref{mechanics}a we show the result as we applied a small sinusoidal cavity length modulation of about 3 nm to probe one flank of the cavity resonance (see inset in Fig.\,\ref{mechanics}b). Figure\,\ref{mechanics}b displays the fast Fourier transform of the signal when the cavity-molecule coupling was modulated at frequencies of 10 Hz and 114 kHz. Operation at modulation frequencies larger than $\gamma$ would result in well-resolved frequency side bands. Such high vibration frequencies are, indeed, within reach if one fabricates the micromirror on cantilevers with very short lengths or uses higher harmonics of their mechanical motion \cite{Nievergelt:14}.

\section{Discussion and Outlook}
The coupling of organic molecules to microcavities has been of interest to different communities \cite{Norris:97, Lidzey:98, Lidzey:00, Steiner:05, Kena-Cohen:08, Chizhik:09, Toninelli:10b, Schwartz:11, Chizhik:12, Shalabney:15, Schmitt:16, Konrad:16}. Some groups have demonstrated the coupling of  single-molecule fluorescence to a cavity mode \cite{Norris:97, Toninelli:10b} while others have investigated fluorescence lifetime changes with single-molecule sensitivity \cite{Steiner:05, Chizhik:09, Chizhik:12, Konrad:16}. However, coherent cavity effects on molecules have only been reached in ensembles \cite{Lidzey:98, Lidzey:00, Kena-Cohen:08, Schwartz:11, Shalabney:15, Schmitt:16} or by exploiting plasmonic nanoparticles \cite{Chikkaraddy:16}. In fact, all coherent studies have been performed at room temperature and thus suffer from severe dephasing of the order of $10^5$ due to the coupling of  the molecular dipole moment to the thermal fluctuations of the matrix. In this work we have presented the first demonstration of coherent interaction between a single organic molecule and a microcavity. The noteworthy and crucial features of our work are the low mode volume and low $Q$ of the cavity \cite{Kelkar:15}. This experimental regime is decisive because it provides access to a large number of molecules within one cavity resonance and avoids the need for stabilizing ultranarrow cavity resonances in a cryostat. In addition, the introduction of an organic crystal does not lower the cavity $Q$ further. 

Our work opens doors to several new directions in molecular quantum optics. First, improving the $Q/V$ ratio by a moderate amount of only one order of magnitude will result in more than 90\% attenuation of a photon in the cavity. Such a regime brings about many linear and nonlinear benefits such as efficient sources of narrow-band single photons or photonic switching. Moreover, an efficient coupling ushers in a new paradigm, where several molecules can coherently couple via the cavity field \cite{Auffeves:11, Galego:15, Haakh:15, Haakh:16} without the need for near-field coupling \cite{Hettich:02}. To facilitate many of these exciting experiments, we plan to replace the silver mirror with dielectric coatings and integrate microelectrodes into our sample in order to tune several molecules to the same resonance. 

\textbf{Acknowledgments}
We thank Bj\"orn Hoffmann, Silke Christiansen, and Anke Dutschke for focused ion beam milling of micromirrors and Maksim Schwab for his contributions to the mechanical components of the cryostat. This work was supported by an Alexander von Humboldt professorship, the Max Planck Society and the European Research Council (Advanced Grant SINGLEION). D. W. and H. K. contributed equally to this work.

\end{document}